\documentclass[epsf,usegraphicx]{mn2e}

\usepackage{xspace}
\usepackage{lscape}

\title[{\sl RATS-Kepler}  -- a deep high cadence survey of the Kepler field]
{{\sl RATS-Kepler} -- a deep high cadence survey of the Kepler field}

\author[]
{Gavin Ramsay$^{1}$, Adam Brooks$^{1,2}$, Pasi Hakala$^{3}$, Thomas Barclay$^{4,5}$, \and
David Garcia-Alvarez$^{6,7,8}$,
Victoria Antoci$^{9}$, Sandra Greiss$^{10}$, Martin Still$^{4,5}$, \and
Danny Steeghs$^{10}$, Boris G\"{a}nsicke$^{10}$, Mark Reynolds$^{11}$ \and\\
$^{1}$Armagh Observatory, College Hill, Armagh, BT61 9DG, UK\\
$^{2}$Mullard Space Science Laboratory, University College
London, Holmbury St. Mary, Dorking, Surrey, RH5 6NT, UK\\
$^{3}$Finnish Centre for Astronomy with ESO, University of Turku, V\"{a}is\"{a}l\"{a}ntie 20, FI-21500 PIIKKI\"{O}, Finland\\
$^{4}$NASA Ames Research Center Institute, Moffett Field, CA 94035, USA\\
$^{5}$Bay Area Environmental Research Institute, Inc., 560 Third St. West, Sonoma, CA 95476, USA\\
$^{6}$Instituto de Astrofísica de Canarias, E-38205 La Laguna, Tenerife, Spain\\
$^{7}$Dpto. de Astrofísica, Universidad de La Laguna, 38206 La Laguna, Tenerife, Spain\\
$^{8}$Grantecan CALP, 38712 Breña Baja, La Palma, Spain\\
$^{9}$Stellar Astrophysics Centre, Dept of Physics and Astronomy, Aarhus University, Ny Munkegade 120, DK-8000 Aarhus C, 
Denmark\\
$^{10}$Department of Physics, University of Warwick, Coventry, CV4 7AL, UK\\
$^{11}$Department of Astronomy, University of Michigan, 500 Church Street, Ann Arbor, MI 48109, USA\\
}

\date{Accepted 2013 October 1.  Received 2013 September 27; in original form 2013 April 2}

\begin{document}
\outer\def\gtae {$\buildrel {\lower3pt\hbox{$>$}} \over {\lower2pt\hbox{$\sim$}} $}
\outer\def\ltae {$\buildrel {\lower3pt\hbox{$<$}} \over {\lower2pt\hbox{$\sim$}} $}
\newcommand{\ergscm} {ergs s$^{-1}$ cm$^{-2}$}
\newcommand{\ergss} {ergs s$^{-1}$}
\newcommand{\ergsd} {ergs s$^{-1}$ $d^{2}_{100}$}
\newcommand{\pcmsq} {cm$^{-2}$}
\newcommand{\ros} {\sl ROSAT}
\newcommand{\kep} {\sl Kepler}
\newcommand{\xmm} {\sl XMM-Newton}
\newcommand{\swift} {\sl Swift}
\def\rchi{{${\chi}_{\nu}^{2}$}}
\newcommand{\Msun} {$M_{\odot}$}
\newcommand{\Mwd} {$M_{wd}$}
\def\Mdot{\hbox{$\dot M$}}
\def\mdot{\hbox{$\dot m$}}
\newcommand{\teff}{\ensuremath{T_{\mathrm{eff}}}\xspace}

\maketitle

\begin{abstract}

We outline the purpose, strategy and first results of a deep, high
cadence, photometric survey of the {\kep} field using the Isaac Newton
Telescope on La Palma and the MDM 1.3m Telescope on Kitt Peak. Our
goal was to identify sources located in the {\kep} field of view which
are variable on a timescale of a few mins to 1 hour. The
astrophysically most interesting sources would then have been
candidates for observation using {\kep} using 1 min sampling. Our
survey covered $\sim$42\% of the {\kep} field of view and we have
obtained light curves for 7.1$\times10^5$ objects in the range
13$<g<20$. We have discovered more than 100 variable sources which
have passed our two stage identification process. As a service to the
wider community, we make our data products and cleaned CCD images
available to download. We obtained {\kep} data of 18 sources which we
found to be variable using our survey and we give an overview of the
currently available data here. These sources include a pulsating DA
white dwarf, eleven $\delta$ Sct stars which have dominant pulsation
periods in the range 24 min to 2.35 hrs, three contact binaries, and a
cataclysmic variable (V363 Lyr). One of the $\delta$ Sct stars is in a
contact binary.

\end{abstract}


\section{Introduction}

The prime objective of the {\sl Kepler} mission is to detect Earth
sized planets orbiting Solar type stars in the habitable zone (Koch et
al 2010). It does this by detecting transits of the host star by the
orbiting exoplanet. The lightcurves which {\kep} obtained extended
over many months and have a precision of parts per million. These data
allow models of stellar structure to be tested in a way that has not
been possible before (e.g. Bedding et al. 2011). Furthermore, it has
led to the unexpected discovery of extreme binary systems such as the
`{\sl Heartbeat}' stars which are excellent tests of binary and
stellar models (Welsh et al. 2011, Thompson et al. 2012).

Asteroseismology provides the means to probe the masses and
compositions of stellar interiors; determine stellar internal rotation
profiles; the extent of instability strips and therefore test models
of stellar structure and evolution (e.g. Chaplin et al. 2011).  To
study compact objects such as pulsating white dwarfs, relatively high
cadence observations are essential.  For the vast majority of
observations made using {\kep}, the effective exposure time is 30 mins
(`{\sl Long Cadence}'). However, for a much more limited number of
stars (512) a shorter effective exposure of 1 min is possible (`{\sl
  Short Cadence}').

Before the launch of {\kep}, an extensive programme to identify bright
G/K dwarfs with minimial stellar activity was carried out. Although a
small number of photometric variability surveys were carried out
pre-launch (e.g. Hartman et al. 2004, Pigulski et al. 2009, Feldmeier
et al 2011) they were either not especially deep, did not have wide
sky coverage or did not have a cadence shorter than a few minutes.  To
fill this gap we started a photometric variability survey ({\sl
  RATS-Kepler}) in the summer of 2011 using the Wide Field Camera on
the Isaac Newton Telescope (INT). Sources which were considered
astrophysically interesting based on their light curve and colour
would then have been the subject of bids to obtain {\kep} Short
Cadence observations.

\section{Photometric Observations}

Our strategy is a modified version of that used by us in the RApid
Temporal Survey ({\sl RATS}) which was carried out using the INT
between 2003 and 2010 (Ramsay \& Hakala 2005, Barclay et al. 2011). In
that project we obtained a series of 30 sec exposures of a given field
in white light for 2.0--2.5 hrs. The resulting lightcurves had a
resulting cadence of $\sim$1 min and, for sources brighter than
$g$=21, the standard deviation ($\sigma$) of the light curves was
$<$0.024 mag (Barclay et al. 2011). It led to the discovery of a rare
double-mode pulsating sdB star (Ramsay et al. 2006, Baran et
al. 2011), pulsating white dwarfs and several dozen distant $\delta$
Sct or SX Phe stars (Ramsay et al. 2011).

Since the {\kep} field of view (116 square degrees) is more than twice
the area covered by the {\sl RATS} project, we decided to increase the
number of fields observed per night by obtaining a one hour (rather
than a two hour) sequence of short exposures per pointing. Since the
photometric precision of {\kep} Short Cadence observations reduces
from 12.9 percent at $g$=19 to 32.4 percent at $g$=20 (this compares
with 0.85 percent at
$g$=16)\footnote{http://keplergo.arc.nasa.gov/CalibrationSN.shtml} we
also reduced the exposure to 20 sec and used the $g$ band filter
instead of white light (since sources fainter than $g$=20 would give
high photometric errors in {\kep} observations).

\begin{figure}
\begin{center}
\setlength{\unitlength}{1cm}
\begin{picture}(8,8)
\put(-0.5,-1){\includegraphics{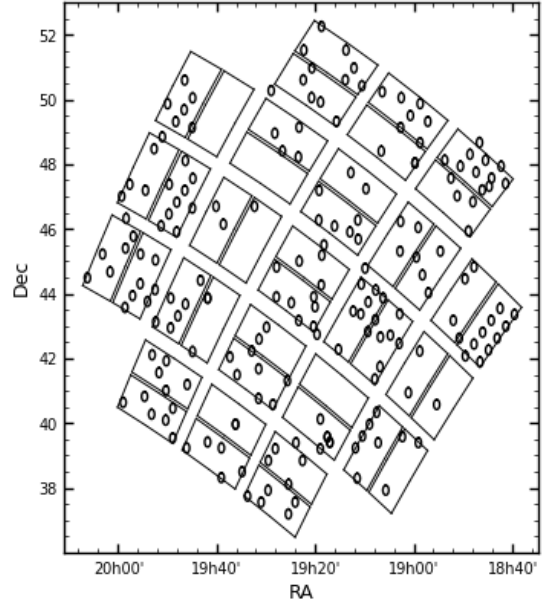}}
\end{picture}
\end{center}
\caption{The field centers of our pointings which have been observed
  to date. Each circle has a diameter of 0.3 square degrees.}
\label{radec}
\end{figure}

During the summer of 2011 and 2012 we obtained data using the 2.5m INT
located on the island of La Palma, and the 1.3m MDM McGraw-Hill
Telescope located on Kitt Peak (see Table 1 for details). Our
observations cover 42 percent of the {\sl Kepler} field.  In Figure
\ref{radec} we show the position of stars observed in our survey in
equatorial coordinates.

\subsection{Isaac Newton Newton Telescope}

The INT Wide Field Camera has 4 CCDs and covers 0.29 degrees squared.
The deadtime was 30 sec, giving a cadence of $\sim$50 sec. We are
sensitive to flux variations on timescales as short as a few mins in
sources with a magnitude in the range 13.5$<g<$21. Tables A1 and A2
show the dates and field centers of each pointing.

\subsection{MDM Telescope}

The red4k detector was used for two nights and the MDM4k detector for
five nights on the MDM
Telescope\footnote{http://mdm.kpno.noao.edu}. The field of view in
both detectors is 0.12 degrees squared. We used a 30 sec exposure in
the $V$ band for our sequence of observations. For the red4k detector,
the readout of 55 sec gave a cadence of $\sim$85 sec while for the
MDM4k detector the readout was 40 sec, giving a cadence of $\sim$70
sec. For each field we obtained an image in $BR$, which were then
transformed into $gr$ magnitudes (Jester et al. 2005) with appropriate
normalisation to the Kepler INT Survey (KIS, Greiss et al. 2012a,b)
results. Table A3 shows the dates and field centers of each pointing.

\begin{table}
\begin{center}
\begin{tabular}{lrrr}
\hline
Dates & Telescope & No. Fields  \\
\hline
11--17 Jul 2011 & INT & 49 \\
01--10 Aug 2011 & INT & 58 \\
16--22 May 2012 & MDM & 26 \\
03--12 Aug 2012 & INT & 55 \\
\hline
\end{tabular}
\end{center}
\caption{The dates of observations made using the 2.5m Issac Newton
  Telescope on La Palma and the 1.3m MDM Telescope on Kitt Peak. We
  note the number of individual fields which were subject to a 1 hr
  sequence of short exposures in the g band (INT) and V band (MDM).}
\end{table}

\subsection{Image Reduction}

\begin{figure*}
\begin{center}
\setlength{\unitlength}{1cm}
\begin{picture}(16,9)
\put(17,-0.8){\includegraphics{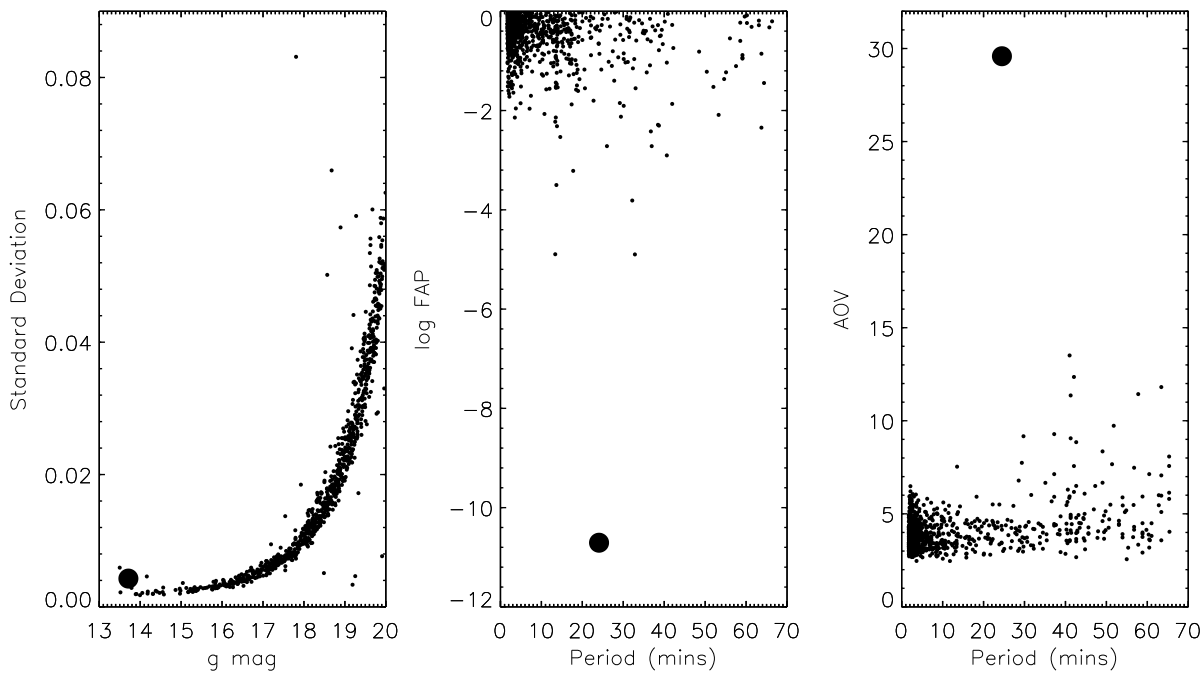}}
\end{picture}
\caption{From left to right we plot for all stars in field 73 chip4
  (which includes the short period blue variable KIC 3223460) the
  $\sigma$ of each light curve as a function of $g$ mag; the log FAP
  as a function of the period of the highest peak in the LS power
  spectrum; and the AoV value as a function of the AoV period.}
\label{compare}
\end{center}
\end{figure*}

The data were corrected for the bias level and were flat-fielded using
{\tt Starlink}\footnote{http://starlink.jach.hawaii.edu/starlink} and
{\tt ftools}\footnote{http://heasarc.gsfc.nasa.gov/ftools} software.
To embed sky coordinates into the images we used software made
available by astrometry.net (Lang et al. 2010). The resulting
astrometric positions agreed with the 2MASS source catalogue typically
within 0.4 arcseconds. We cross-correlated the positions of our
sources with that of the KIS (Greiss et al. 2012a,b) which reaches
down to a limit of 20th mag in $U, g, r, i$ and H$\alpha$ filters. We
also obtained a single $r$ band image of every INT field and a single
image of every MDM field in the $BR$ filters. We corrected our
instrumental magnitudes by scaling them to agree with the KIS values
using an offset derived for matched sources.

\section{Data Analysis}

We broadly follow the same data reduction and analysis strategy as we
used for the {\sl RATS} project, which is described in Barclay et al.
(2011). However, we now outline some features specific to the {\sl
  RATS-Kepler} project.

\subsection{Extracting light curves}

The {\kep} field extends 6--21 degrees above the Galactic plane. Each
of our individual fields are therefore relatively crowded at low
latitudes or surprisingly sparse at higher latitudes. For sparse
fields we used {\tt sextractor} (Bertin \& Arnouts 1996) to extract
magnitudes using aperture photometry. Differential magnitudes were
determined by comparing the magnitude of each star with the mean
brightness of the 3--10th most brightest stars in the image (the
results were very similar if we chose, say, the 4--20th most brightest
stars).  For more crowded fields we used {\tt diapl2}, an updated
version of {\tt diapl} (Wozniak 2000), which extracts photometry by
applying the well established `Difference Imaging Subtraction' method
(Alard \& Lupton 1998).

Despite the fact that differential photometry has been performed we
find that the photometry of certain fields suffer from systematic
trends in the data -- ie the light curves derived from the same
detector can show similar features. This effect can be seen in many
other large scale surveys, including {\kep} (Kinemuchi et al 2012). In
our case, the systematic trends will largely result from the fact that
for good technical reasons we do not use the autoguider for our INT
observations. To ensure that stars remain roughly at the same position
on the detector we apply manual corrections to the pointing.

However, we aimed to remove the effects of systematic trends by
appling the {\tt SYSREM} algorithm (Tamuz et al. 2005) to the light
curves derived from each CCD individually using a varying number of
cycles as described in Tamuz (2006) (see also Barclay et al. 2011).
For a small number of fields it was not possible to detrend the data
(mainly because of the low number of stars available). Using a faint
limit of $g=20.0$ and a bright limit of $g=13.5$, we have obtained a
total of 7.1$\times10^5$ detrended light curves.

\begin{figure}
\begin{center}
\setlength{\unitlength}{1cm}
\begin{picture}(8,6.5)
\put(8.5,0){\includegraphics{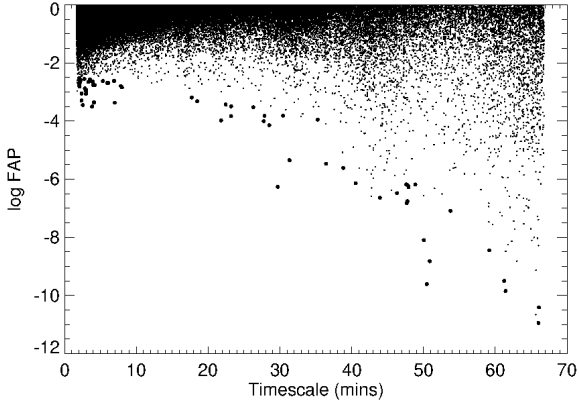}}
\end{picture}
\caption{For those 10$^{5}$ stars located in field-chip combinations
  which have the least negative median log FAP, we plot the log of the
  False Alarm Probability (FAP) as a function of the period of the
  most prominent peak in the Lomb Scargle power spectrum. More
  negative values of log FAP imply a greater chance of intrinsic
  variability. Those sources which have been identified as variable
  candidates using the MAD statistic with $n$=14 are shown as larger
  filled circles (see \S 3.2 for details).}
\label{per-fap}
\end{center}
\end{figure}

\subsection{Identifying variable candidates}

Identifying {\it bona fide} variable stars from a large sample of
light curves is not a trivial task. Here we use a two stage process of
identifying variable stars. In the first stage we use different
statistical tests to obtain a sample of candidate variables. In the
second stage we manually perform a quality assessement of each light
curve and associated images to remove sources which have been
spuriously identified as variable.

Different statistical tests are better suited to identifying different
kinds of variability.  For instance, the Lomb Scargle (LS) Periodogram
(Lomb 1976, Scargle 1982) is particularly well suited to detecting
pulsating variables where the pulsation period is shorter than the
duration of the light curve. On the other hand, the Alarm test (Tamuz,
Mazeh \& North 2006) and the Analysis of Variance (AoV) test
(Schwarzenberg-Czerny 1989, 1996 and Devor 2005 for the implementation
used by {\tt VARTOOLS}) are suitable for identifying eclipsing or
contact binaries, whilst the $\chi^{2}$ test (where the model is the
mean magnitude of the light curve) is good for detecting flare stars
(see Graham et al. 2013 for a recent review of which tests are best
suited to identifying specific kinds of variable star).

The {\tt VARTOOLS} suite of software (Hartman et al. 2008) allows the
parameterisation of large numbers of light curves using many different
statistical tests in a quick and simple manner. We apply the following
tests on each of our light curves: the LS Periodogram; the Alarm test
and the AoV test. We also determine the $\chi^{2}$ value and standard
deviation ($\sigma$) for each light curve after applying a 5 $\sigma$
clipping to each light curve. A file containing the positions and
colours of all sources along with their photometric variability
parameters can be downloaded via Armagh Observatory Web Site
(star.arm.ac.uk/rats-kepler). Table \ref{fits-files-parameters}
outlines the full set of parameters which are given in this FITS file.

As an example of how different tests compare, we show in Figure
\ref{compare} the results of the $\sigma$, LS and AoV tests on a field
which contains KIC 3223460 which has a dominant pulsation period of
24.2 mins (see Table \ref{kepler-sources-table}). Plotting the
$\sigma$ of each light curve as a function of $g$ mag shows that KIC
3223460 has a greater $\sigma$ than the main distribution of sources
with similar magnitude. However, there is (naturally) no information
on the timescale of variability. On the other hand, the LS and AoV
test clearly identify KIC 3223460 as being strongly variable on a
period of 24 mins.

The main goal of our survey is to identify compact pulsating stars in
the {\kep} field which would then have been the subject of bids to
observe them in Short Cadence mode using {\kep}. As demonstrated in
Figure \ref{compare}, the LS Periodogram is efficient at identifying
these sources in our data. The LS Periodiogram as implemented in {\tt
  VARTOOLS} determines the frequency of the highest peak in the power
spectrum (which we define as the `Period' of variability even if the
source cannot be verified as stricty periodic) and the False Alarm
Probability of this peak being statistically significant. For each
light curve we obtained an LS power spectrum and performed the AoV
test in the frequency interval corresponding to the Nyquist frequency
(847.1 cycles/day -- which equates to a period of 1.7 mins -- for the
INT data and 600 cycles/day -- which equates to a period of 2.4 mins
-- for the MDM data) and 21.49 cycles/day -- which equates to a period
of 67 mins -- (which is the mean duration of the INT light curves).

For the purposes of selecting an initial sample of candidate
variables, we use the LS test. In the absence of red noise and
systematic trends, a peak in the power spectrum with log FAP=--2.5 is
likely to be significantly variable at the 3$\sigma$ confidence
level. However, since the seeing and sky brightness can vary from
field to field, and the success of the detrending alogrithim can vary
from chip to chip, the threshold for identifying variables can be more
negative than log FAP = --2.5.

We determined the median value of the log FAP statistic for the light
curves in each field-chip combination. We then ordered our sources by
this median log FAP and made seven sub-sets containing 10$^{5}$ stars
each (the remaining 8787 stars which were in field-chip combinations
with the highest median log FAP were discarded). For the subset with
the least negative mean value for log FAP (Figure \ref{per-fap}),
there were 911 sources which has a log FAP $<$-2.5 (or 0.91 percent of
sources in the sub-set).

Rather than using a fixed threshold for the log FAP to provide an
initial selection of candidate variables, we used the Median Absolute
Deviation (MAD) to provide a means of identifying sources which were
'outliers' in the Period - log FAP distribution. The MAD is defined
for a batch of parameters $\{x_{1},\ldots,x_{m}\}$ as

\begin{equation}
 \mathrm{MAD} = \mathrm{median}_{i} ( | X_{i} - \mathrm{median}_{j} (X_{j}) | )
\end{equation}

We ordered the data by Period and then into 2 min time bin intervals
and derived the MAD for each bin. Candidate variables are then
selected so that variable sources obey (log FAP) $<$ MAD$_{log
  FAP}\times$n + Median$_{log FAP}$, where $n$ is an integer which
defines how far a source is from the local median log FAP. It is
selected largely by trial and error -- too high a value of $n$ will
select only the most strongly variable sources, but too low a value of
$n$ will produce large quantities of candidate variables, all of which
require manual verification. (To be selected as a candidate variable,
a source also has to have log FAP $<$-2.5). The selection of variables
using the MAD statistic was done on each subset of 10$^{5}$ stars
separately then combined according to the $n$ value that was used.  We
found that for $n$=18, 227 stars (or 0.032 percent of the total) were
selected as candidate variables ($n$=16 $\rightarrow$ 368 stars,
$n$=14 $\rightarrow$ 642, $n$=12 $\rightarrow$ 1187, $n$=10
$\rightarrow$ 1999). We stress that this selection simply identifies
those stars which are most likely to be variable.

\begin{figure}
\begin{center}
\setlength{\unitlength}{1cm}
\begin{picture}(16,6)
\put(-0.8,0.){\includegraphics{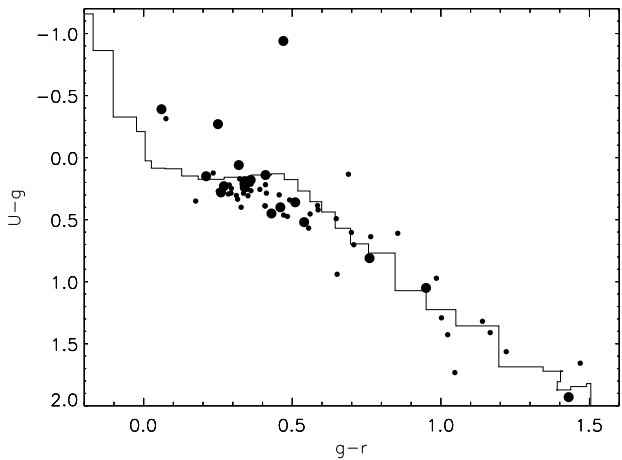}}
\end{picture}
\caption{The solid line shows the unreddended main sequence (taken
  from Groot et al. 2009). Small circles indicate the colours of
  sources shown in Table \ref{variables-list} while larger circles
  indicate the colours of those sources which we have obtained {\kep}
  Short Cadence data (Table \ref{kepler-sources-table}).}
\label{intvariablecolours}
\end{center}
\end{figure}

Each source is then subject to a manual inspection of the light curve
and corresponding images to verify their variable nature. Where
appropriate, additional light curves were obtained using the optimal
aperture photometry routine {\tt autophotom} (Eaton, Draper \& Allan
2009). In this process we assign a Flag value (see Table
\ref{fits-files-parameters}) to characterise their lightcurve and
timescale of variability. For our $n$=18 sample we found that after a
second stage verification process we had 65 sources which we classed
as highly likely to be variable using the LS statistic as a first
screening stage. A summary of these sources are shown in Table
\ref{variables-list} and their colours are shown in Figure
\ref{intvariablecolours}. By examining samples derived using $n$=14,
we have found more than 100 sources which have passed our two stage
variable identification process. These are indicated in the FITS file
which we make available on the Armagh Observatory Web site.

A high fraction of sources (71 percent) of the $n$=18 sample were
found to be unlikely to be {\sl bona fide} variable sources. Many of
these spurious variables were caused by residual systematic trends in
the light curves - for instance sources from the same chip showed very
similar trends in their light curves. This is almost certainly a
result of using manual corrections in the guiding process in the INT
observations and for the fact that the light curves only covered a
short time (typically 1 hr).

\begin{figure*}
\begin{center}
\setlength{\unitlength}{1cm}
\begin{picture}(16,11)
\put(0,0){\includegraphics{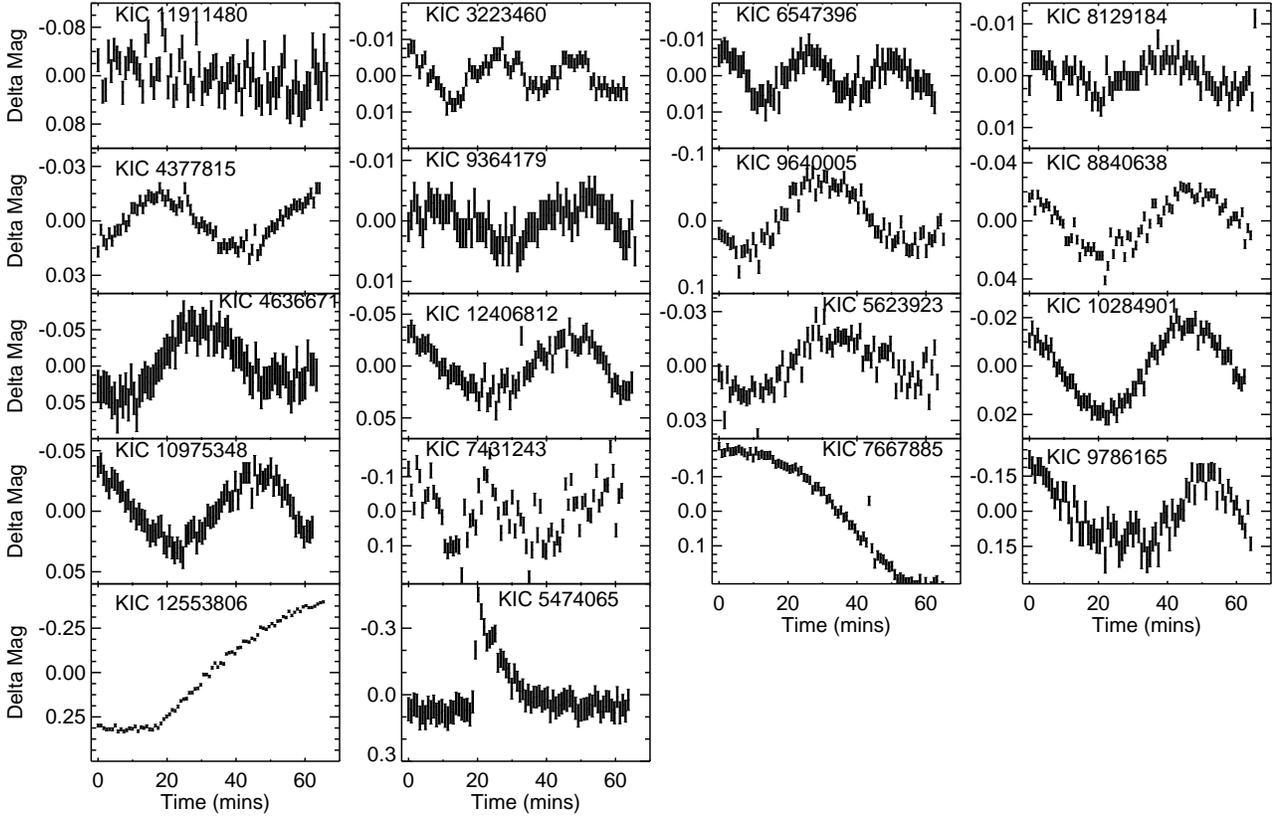}}
\end{picture}
\caption{The INT light curve of those sources which have been
  identified as variable in the {\sl Kepler-RATS} survey and which we
  have obtained {\kep} Short Cadence data. The details of these
  sources are shown in Table \ref{kepler-sources-table}.}
\label{int-light}
\end{center}
\end{figure*}

We selected 18 sources which would be good targets to observe using
{\kep} in 1 min sampling mode. We successfully bid for {\kep} Guest
Observer Programme and the Directors Discretionary Time Programme. We
show the INT light curves of these sources in Figure \ref{int-light}
and we give their sky coordinates, magnitude, colours in Table
\ref{kepler-sources-table}. We indicate an approximate spectral
classification using our INT and Gran Telescopio Canarias (GTC)
spectra (\S \ref{spectra}). Of these 18 sources, one is a pulsating DA
white dwarf (the details are presented in Greiss et al in prep), three
are contact binaries, one is a cataclysmic variable, and one is a
flare star (Ramsay et al. 2013). However, most of these sources appear
to be $\delta$ Sct stars. We will present an overview of the {\kep}
data of these sources in \S \ref{keplerobs}.

\begin{table*}
\begin{center}
\begin{tabular}{rrrrrrlrrrr}
\hline
KIC & RA & DEC & g & U-g & g-r & {\kep} SC & Period & Spectra & Spectral& Variable\\
    & (J2000) & (J2000) & &  & & Data & Kepler & & Type & Type\\
\hline
11911480 & 19 20 24.9 & +50 17 22.4 & 18.13 & --0.39 & 0.06 & 12,16 & 290 sec &  & & DAV (1)\\
3223460  & 19 12 32.2 & +38 23 00.1 & 13.74 & --0.27 & 0.25 & 14&  24.2 min & GTC & mid-late A  & $\delta$ Sct\\
6547396  & 19 53 18.3 & +41 58 26.9 & 14.84 &   0.40 & 0.46 & 16& 26.6 min & INT & mid-late A & $\delta$ Sct\\
8120184  & 19 54 11.9 & +43 59 20.1 & 14.27 &   0.36 & 0.51 & 15& 42.6 min & INT& mid-late A  & $\delta$ Sct\\ 
4377815  & 19 39 08.1 & +39 27 35.9 & 14.83 &   0.24 & 0.34 & 15& 45.7 min & INT& mid-late A  & $\delta$ Sct\\ 
9364179  & 19 56 24.5 & +45 48 24.1 & 14.38 &   0.45 & 0.43 & 15& 46.8 min & INT& mid-late A  & $\delta$ Sct\\ 
9640005  & 19 09 46.3 & +46 20 04.1 & 18.40 &   0.15 & 0.21 & 14--16& 49.5 min & GTC & mid A  & $\delta$ Sct\\
8840638  & 19 55 35.1 & +45 04 46.0 & 14.63 &   0.52 & 0.54 & 14--16& 49.6 min & GTC & mid-late A & $\delta$ Sct\\
4636671  & 19 01 52.2 & +39 45 59.3 & 15.67 &   0.28 & 0.26 & 14--16& 50.0 min & GTC & mid A  & $\delta$ Sct\\
12406812 & 19 23 33.8 & +51 17 58.9 & 17.24 &   0.18 & 0.36 & 14--16&50.4 min & GTC &  mid-late A  & $\delta$ Sct\\ 
5623923  & 19 32 01.5 & +40 51 16.8 & 16.62 &   0.23 & 0.27 & 14--16&50.5 min & GTC & mid-late A  & EB+$\delta$ Scuti \\
10284901 & 19 43 46.4 & +47 20 32.8 & 15.73 &   0.06 & 0.32 & 14,16& 75.8 min & GTC  &  mid-late A  & $\delta$ Sct\\
10975348 & 19 26 46.1 & +48 25 30.8 & 18.89 &   0.19 & 0.34 & 14--16& 2.35 hrs& GTC & mid A  & $\delta$ Sct\\
7431243  & 19 08 51.6 & +43 00 31.5 & 19.10 & --0.94 & 0.47 & 16 & 4.68 hrs &  &  & CV (2)\\ 
7667885  & 19 03 30.2 & +43 23 22.7 & 17.64 &   1.05 & 0.95 & 14--16&7.56 hrs &GTC& mid G & W UMa \\ 
9786165  & 19 50 11.0 & +46 34 40.8 & 17.67 &   0.81 & 0.76 & 14,16& 7.98 hrs &GTC & mid G & W UMa \\  
12553806 & 19 14 41.0 & +51 31 08.9 & 17.52 &   0.14 & 0.41 & 14--16&11.12 hrs &GTC &  A+F? & W UMa\\ 
5474065  & 19 53 02.5 & +40 40 34.6 & 18.77 &   1.93 & 1.43 & 14&Flare &GTC & M3 V & Flare star (3)\\ 
\hline
\end{tabular}
\caption{The details of those sources whose light curve is shown in
  Figure \ref{int-light}. We indicate their KIC ID; their RA and Dec;
  the $g$ mag, $(U-g)$ and $(g-r)$ colours, which have been taken from
  the KIS (Greiss et al. 2012a,b). We also show in which Quarter
  {\kep} Short Cadence (SC) data was obtained and what the dominant
  period was in the power spectrum of the {\kep} light curve. In the
  'Spectra' column we indicate if we obtained a spectrum of the source
  using the INT or GTC, what the approximate spectral type was and in
  the last column what the type of variable star the source is. EB:
  Eclipsing Binary. Notes: (1) Greiss et al. in prep, (2) Cataclysmic
  Variable, Scaringi et al. 2013), (3) Ramsay et al. (2013).}
\label{kepler-sources-table}
\end{center}
\end{table*}

\section{Optical Spectroscopy}
\label{spectra}

As part of our follow up programme, we obtained low--medium resolution
optical spectroscopy of over 50 sources which were either identified
as being variable on a short timescale or had unusual colours.  (One
very blue source in our sample, KIC 10449976, has already been
reported as an extreme helium star, Jeffery et al. 2013).  We obtained
data using the Intermediate Dispersion Spectrograph (IDS) and the R400
grism on the INT between 26--28 June 2012 and also using the Optical
System for Imaging and Low Resolution Integrated Spectroscopy (OSIRIS)
tunable imager and spectrograph and the R1000R grism on the 10.4\,m
GTC during March -- June 2013. Both are located at the Observatorio
Roque de los Muchachos in La Palma, Canary Islands, Spain.  At least
two spectra were obtained of each source and the individual exposure
time ranged from 180 to 360 sec (INT) and 40 to 400 sec (GTC). The
spectra were reduced using standard procedures with the wavelength
calibration being made using a CuNe+CuAr arc taken shortly after the
object spectrum was taken. A flux standard was observed so that the
(resulting combined) spectra of each source could be flux calibrated
in the case of the IDS spectra and to remove the instrumental response
in the OSIRIS spectra (the observing programme utilises poor observing
conditions). The spectral resolution of our INT spectra was
$\sim$2\AA\hspace{1mm} and $\sim$8\AA\hspace{1mm} for our GTC spectra.

\begin{figure}
\begin{center}
\setlength{\unitlength}{1cm}
\begin{picture}(8,9)
\put(0.0,-0.8){\includegraphics{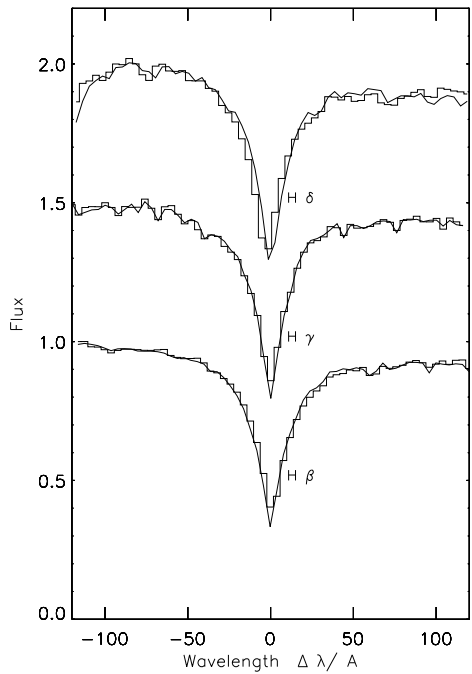}}
\end{picture}
\caption{As an example of the spectral fits to our optical spectra, we
  show the fit to KIC 3223460. For each spectral line, the continuum has been 
normalised to unity and each spectrum after H$_{\beta}$ has been shifted up by 
0.5 flux units.}
\label{spec-balmer-fits}
\end{center}
\end{figure}

For stars which were of the spectral type A/F we modelled the spectra
using a grid of LTE models calculated using the {\sc atlas9} code
(Kurucz 1992) with convective overshooting switched off. Spectra were
calculated with the {\sc linfor} line-formation code (Lemke
1991). Data for atomic and molecular transitions were compiled from
the Kurucz line list.  The stellar temperatures were estimated from
the hydrogen Balmer lines of the stars (H$\beta$ to H$\delta$) using
the {\sc fitsb2} routine (Napiwotzki et al. 2004). No gravity
sensitive features are accessible in our low resolution spectra.
McNamara (1997) finds that SX Phe and large amplitude $\delta$ Sct
stars have a range in log g of 3.0--4.3. In our fits we fixed log
g=4.0, although the resulting temperature is only weakly sensitive to
this parameter. The metallicity was allowed to vary although this was
not strongly constrained in the fits. The error of the fit parameters
were determined with a bootstrapping method. As an example of the fits
we show in Figure \ref{spec-balmer-fits} the fit to the spectrum of
KIC 3223460.  We show in Table \ref{spectral-fits} the temperature we
derive for the stars which we have obtained {\kep} Short Cadence data
(\S \ref{keplerobs}). We make all the spectra available through the
the Armagh Observatory Web site (http://star.arm.ac.uk/rats-kepler)
together with the fitted temperature of each.

\begin{table}
\begin{center}
\begin{tabular}{lrrrr}
\hline
KIC & T$_{GTC}$ & T$_{IDS}$ & T$_{KIC}$ & log g$_{KIC}$  \\
\hline
3223460 & 8180$\pm$110 &       & 7930 & 4.1 \\
6547396 &      & 8290$\pm$110  & 7480 & 4.0 \\
8120184 &      & 7760$\pm$180  & 7290 & 3.7 \\
4377815 &      & 7880$\pm$160  & 7740 & 4.0 \\
9364179 &      & 7860$\pm$180  & 7000 & 4.0 \\
9640005 & 7730$\pm$180 &       &      &       \\
8840638 & 7860$\pm$120 &       & 6310 & 3.8\\
4636671  & 7950$\pm$120 &      & 7640 & 4.0\\          
12406812 & 7660$\pm$140 &      & 7370 & 4.1\\
5623923  & 7970$\pm$110 &      & 8300 & 3.9\\
10284901 & 7710$\pm$180 &      & 8420 & 4.0\\ 
\hline
\end{tabular}
\caption{For those sources which we have observed using {\kep} we show
  the temperature we derive from INT and GTC spectra along with the
  temperature and log g taken from the {\kep} Input Catalog (Brown et
  al. 2011). The errors for T$_{GTC}$ and T$_{INT}$ refer to the
  formal 3$\sigma$ confidence interval. Including systematic
  uncertainties, which have not been included, we estimate the
  realistic 3$\sigma$ uncertainties are $\pm$300 K. The uncertainties
  for T$_{KIC}$ and log g$_{KIC}$ are $\pm$250 K and $\pm$0.25 dex
  respectively (c.f. Brown et al. (2011) and Pinsonneault et
  al. (2012)).}
\label{spectral-fits}
\end{center}
\end{table}

\section{{\kep} observations}
\label{keplerobs}

The detector on board {\kep} is a shutterless photometer using 6 sec
integrations and a 0.5 sec readout. There are two modes of
observation: {\it long cadence} (LC), where 270 integrations are
summed for an effective 28.4 min exposure, and {\it short cadence}
(SC), where 9 integrations are summed for an effective 58.8 sec
exposure.  When an object is observed in SC mode, LC data is also
automatically recorded. After the data are corrected for bias,
shutterless readout smear and sky background, light curves are
extracted using simple aperture photometry (SAP).  Data which were
contaminated, for instance during intervals of enhanced solar
activity, were removed by requiring data to be flagged by the {\tt
  FITS} keyword `SAP\_QUALITY'=0, and the data were corrected for
systematic trends (Kinemuchi et al. 2012).

The {\kep} data on the pulsating DA white dwarf will be presented in
full by Greiss et al. (in prep) while the {\kep} data on the flare star
KIC 5474065 has been presented by Ramsay et al. (2013). Here we give a
brief overview of the {\kep} data on the pulsating, contact binaries
and cataclysmic variable which we have obtained.

\subsection{$\delta$ Sct stars}

\begin{figure*}
\begin{center}
\setlength{\unitlength}{1cm}
\begin{picture}(16,23)
\put(-0.4,-1){\includegraphics{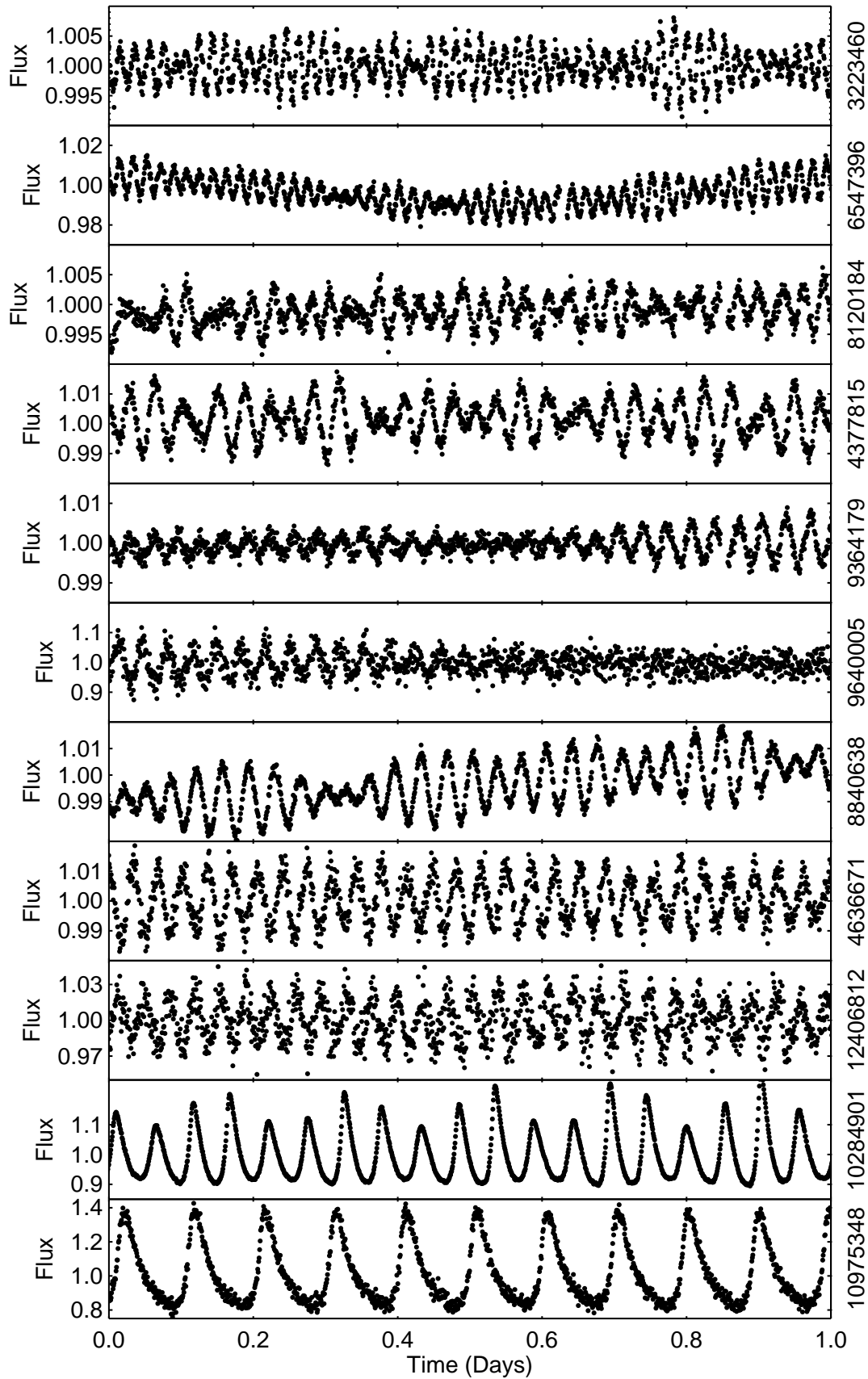}}
\end{picture}
\caption{The {\kep} SC light curves of the eleven sources which appear
  to be $\delta$ Sct type stars. They are ordered so that those
  showing the shortest dominant period is at the top of the figure.
  For clarity we show the light curve covering only one day for each
  star. The KIC identifier for each star is shown on the right hand
  edge of each panel.}
\label{kepler-light}
\end{center}
\end{figure*}

\begin{figure*}
\begin{center}
\setlength{\unitlength}{1cm}
\begin{picture}(16,22)
\put(0,-0.6){\includegraphics{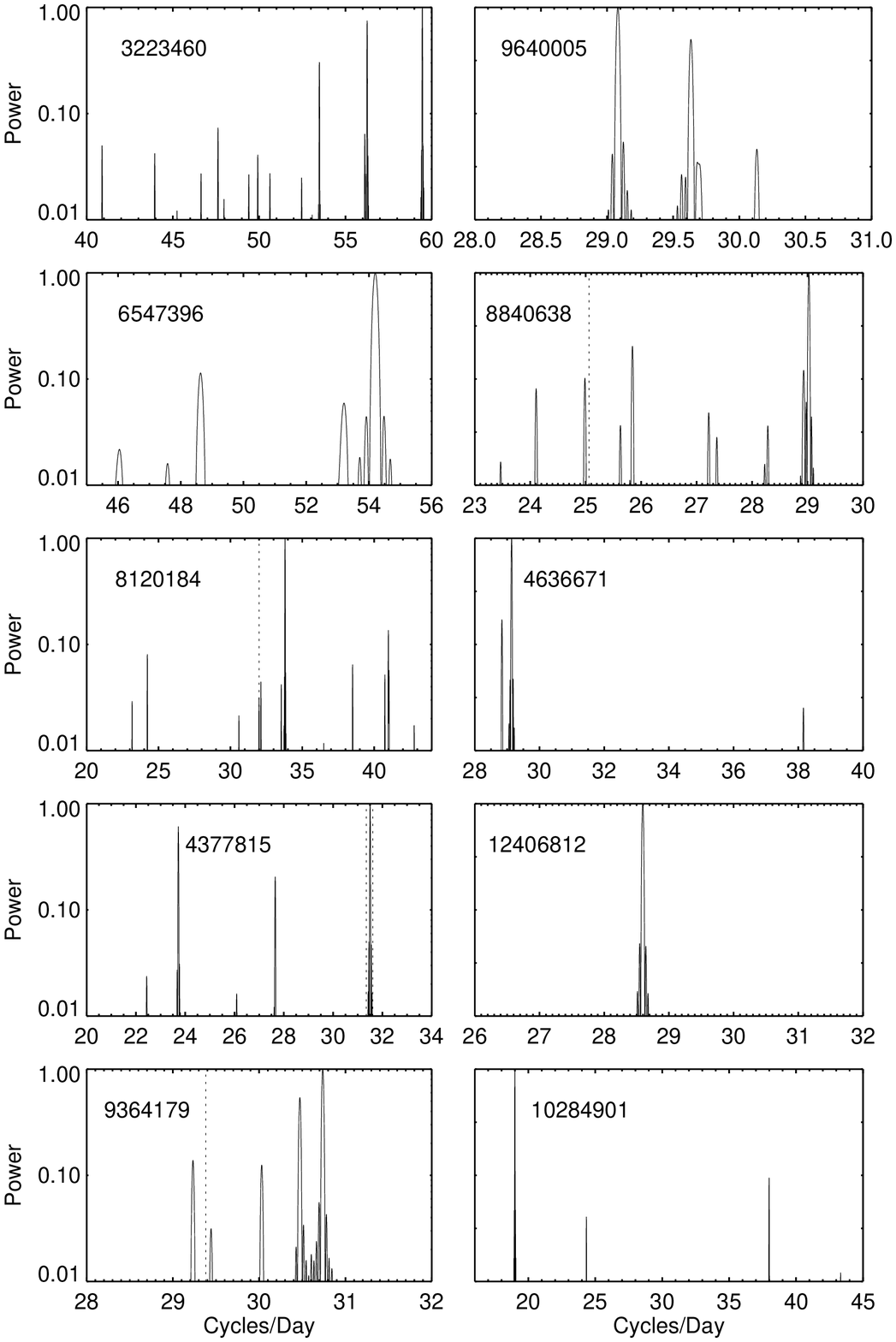}}
\end{picture}
\caption{The power spectra of ten of the sources shown in Figure
  \ref{kepler-light} (we omit for reasons of space the longest period
  system shown in Figure \ref{kepler-light}). We have normalised the
  power spectra so that maximum power is unity, plot the power in log
  space, and we focus on the frequency range of interest in each
  star. The KIC identifier for each star is shown in each panel. The
  light curves for each star shown here cover one month of data except
  for KIC 6547396 the data was taken in quarter 16 and covers 5.2
  days. Dashed vertical lines indicate known artifacts in the power
  spectra of {\kep} SC data.}
\label{kepler-power}
\end{center}
\end{figure*}

In Table \ref{kepler-sources-table} we identify eleven sources which
show a dominant period in the range 24.2 min to 2.35 hrs. Our fits to
their low resolution spectra (Table \ref{spectral-fits}) indicate that
they have a temperature in the range $\sim$7600--8300K. They are
therefore consistent with the characteristics of $\delta$ Sct and SX
Phe stars (see Breger 2000 for a review). We show a short section (1
day) of each of the {\kep} light curves of these $\delta$ Sct type
stars in Figure \ref{kepler-light}. Using a full month of data, we
find that their power spectra are complex and show many frequencies
(Figure \ref{kepler-power}).

The $\delta$ Sct stars with the shortest pulsation periods in our
sample are KIC 3223460 (24 min) and KIC 6547396 (26 min). Indeed, they
are at the extreme short period end of the $\delta$ Sct star
distribution (18 mins marks the short period end, Uytterhoeven et
al. 2011). The discovery of two $\delta$ Sct stars with such a short
period will therefore provide an opportunity to discover the internal
structure of these sources through asteroseismology. Eight of our
sample, (KIC 8120184, 4377815, 9364179, 9640005, 8840638, 4636671 and
5623923), have a peak in their power spectra which lies in the range
42.6 -- 50.5 mins. They have a best fit temperature in the range
7660--7950 K (Table \ref{spectral-fits}). The longest period
pulsators, KIC 10284901 (75.8 mins) and 10975348 (2.35 hrs) also show
high amplitude variations and appear to be high amplitude $\delta$
scuti stars which occupy a restricted range of the instability strip
(McNamara 2000).

Decades of research have shown that the light curves of $\delta$ Sct
stars are very complex (e.g. Pamyatnykh 2000, Breger 2000). However,
this makes them very useful astrophysical laboratories as they show
physical phenomena which can be used to test theoretical
models. However, in many $\delta$ Sct stars it is difficult to
uniquely identify modes in the power spectra of the light curve. An
exception is slow rotators such as 44 Tau (Lenz et al. 2010) where a
large variety of pressure and gravity modes were identified. For the
majority of these stars, when it comes to modelling their power
spectra, a major difficulty is that the mechanism selecting which
modes are excited to observable amplitudes is not well understood
(Dziembowski \& Krolikowska 1990). In other words, some modes are
excited while others are not, which makes identifying the specific
mode for each peak in the frequency spectra difficult.

The $\delta$ Sct stars are stars with masses between 1.5 and 2.5
\Msun \hspace{1mm} and their pulsations are thought to be driven
largely by the opacity mechanism in the He{\sc ii} ionisation zone
(Baker \& Kippenhahn 1962). From {\kep} and {\sl CoRoT} observations,
however, it seems that the opacity mechanism alone cannot excite the
entire range of observed modes. This means that either the models are
incomplete or that there is an additional mechanism contributing to
the driving. Such an alternative explanation would be the presence of
stochastically excited modes, like in the Sun. Theoretical models in
fact predict the the convective envelopes of $\delta$ Sct stars are
still deep and effective enough to drive Solar-like
oscillations. Recently, Antoci et al.  (2011), suggested the detection
of such a hybrid star, showing $\kappa$ mechanism and stochastically
driven modes. However longer observations revealed that the
interpretation is more complicated than initially anticipated (Antoci
et al. 2013). It is therefore of great importance to find pulsating
stars with similar temperature and gravity to HD 187547
($T_{eff}$=7500$\pm$250 K, log g =3.9$\pm$0.25, Antoci et al. 2011) to
test current models. Many of our $\delta$ Sct stars have a similar
temperature to HD 187547 but spectra with higher spectral resolution
than the ones we present in \S 4 are required to provide a robust
temperature determination. However, even if our $\delta$ Sct stars are
likely to be too faint to identify their pulsation modes (even the
brightest of our sources at $g$=13.7 is relatively faint for such an
analysis), the frequency range and the stability of excited modes can
lead to a better understanding of the pulsation mechanisms, provided
the temperature is robustly determined.

\subsection{Contact Binaries}

\begin{figure*}
\begin{center}
\setlength{\unitlength}{1cm}
\begin{picture}(18,8)
\put(0,-2.5){\includegraphics{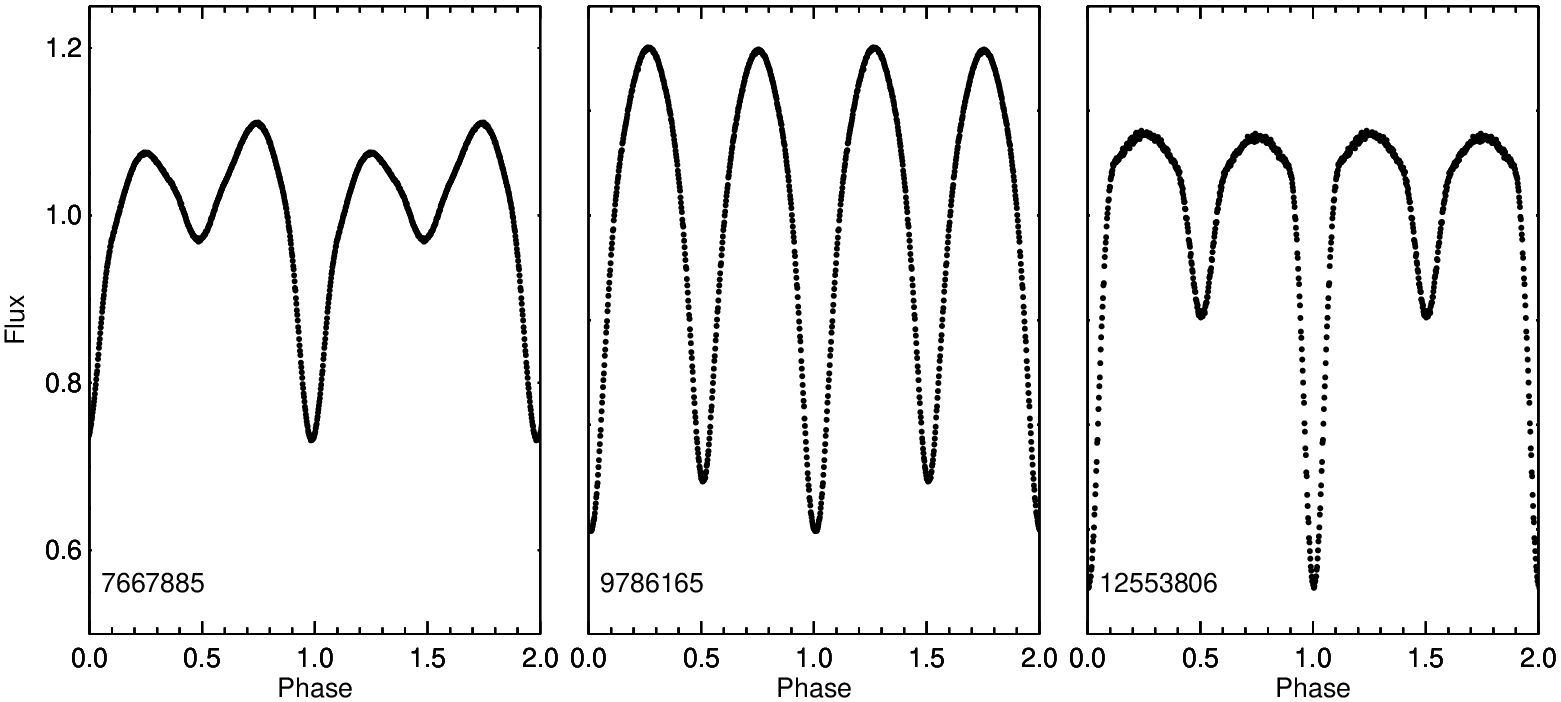}}
\end{picture}
\caption{The {\kep} light curve of the contact binaries KIC 7667885,
  KIC 9786165 and KIC 12553806. The data have been phased so that the
  primary eclipse is centered on $\phi$=0.0 and the y-axis is plotted
  on the same scale in each panel.}
\label{eb-light}
\end{center}
\end{figure*}

The {\kep} data of three of the sources shown in Table
\ref{kepler-sources-table} and Figure \ref{int-light} clearly indicate
that they are eclipsing or contact binaries with an orbital period
ranging from 0.315 days -- 0.463 days. We show the {\kep} Q14 SC data
of these sources in Figure \ref{eb-light} where we have folded and
binned the data on the orbital period.

Pr\u{s}a et al. (2011) and Slawson et al. (2011) present an analysis of
the first and second {\kep} data release of 4044 eclipsing and contact
binaries. Although the shape of the folded light curves of KIC 7667885
and KIC 12553806 are similar to that of semi-detached binaries (also
known as $\beta$ Lyr binaries), their relatively short orbital period
suggests that they are more likely to be contact binaries (also known
as W UMa binaries). The folded light curve of KIC 9786165 also implies
it is a contact binary. Although some caution has to be applied in
interpreting the results of our spectral fits since we have applied a
single temperature model to sources which are clearly binary systems,
the temperatures which we derive (Table \ref{spectral-fits}) indicate
they are contact binaries rather than semi-detached binaries which
have B star components (and hence much hotter).

Unlike the three sources outlined here which have been observed using
SC mode, none of the sources shown in Pr\u{s}a et al. (2011) or
Slawson et al. (2011) have been observed in SC Mode. Since each of the
three binaries have a high inclination, they are excellent data-sets
to search for third bodies (such as exo-planets) in these systems.

\subsection{KIC 5623923: A $\delta$ Sct star in a contact Binary}

\begin{figure*}
\begin{center}
\setlength{\unitlength}{1cm}
\begin{picture}(18,4)
\put(0.8,-5.5){\includegraphics{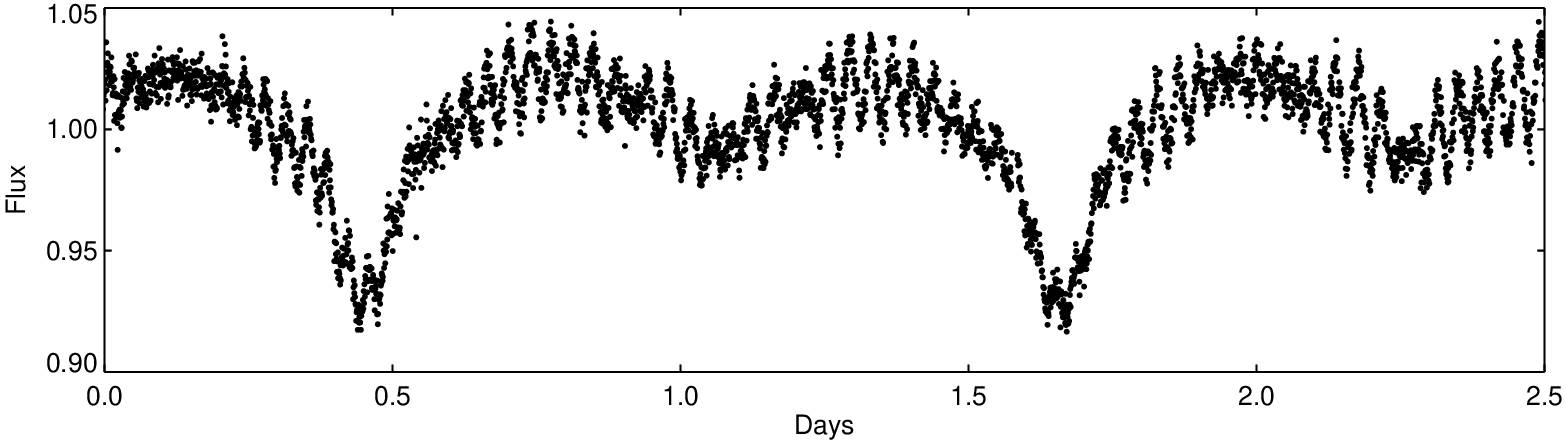}}
\end{picture}
\caption{A short section of Q14 {\kep} data of KIC 5623923. The binary
  component which is obscured during the secondary eclipse shows clear
  evidence of pulsations.}
\label{delta-eb}
\end{center}
\end{figure*}

We show a 2.5 day section of the light curve of KIC 5623923 in Figure
\ref{delta-eb}. It is clear that this source is an eclipsing or
contact binary system with a orbital period of 1.21 days. However,
there are clear pulsations on a period of $\sim$50 mins superimposed
on the light curve. The pulsations are not readily apparent during the
secondary eclipses indicating that the secondary star (the less
luminous of the binary components) is the source of the pulsations.

There are at least two other eclipsing or contact binary stars in the
{\kep} field which have a $\delta$ Sct component. KIC 4544587 is a
binary system with a 2.18 day orbital period (Hambleton et al. 2013)
while KIC 10661783 has an orbital period of 1.23 days (Southworth et
al 2011).  Unlike KIC 5623923 where the secondary star is the
pulsating component, in both KIC 4544587 and KIC 10661783 the primary
is the pulsating star. We note that the spectral fits suggest a
temperature of 8000 K (Table \ref{spectral-fits}) although we have
fitted only a single temperature to a system which is clearly a
binary.

\begin{figure}
\begin{center}
\setlength{\unitlength}{1cm}
\begin{picture}(18,5.5)
\put(0,0){\includegraphics{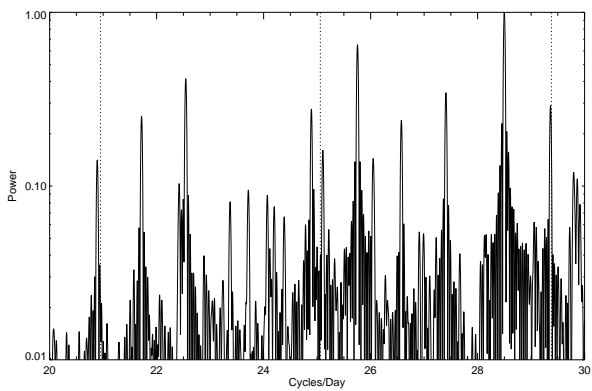}}
\end{picture}
\caption{The power spectrum of the Q14 {\kep} light curve of KIC
  5623823. The spacing between some peaks correspond to the orbital
  frequency suggesting that the amplitude of the $\delta$ Sct
  pulsations are correlated with the orbital period. Dashed vertical
  lines indicate known artifacts in the power spectra of {\kep} SC
  data.}
\label{delta-eb-power}
\end{center}
\end{figure}

The power spectrum of KIC 5623923 (Figure \ref{delta-eb-power}) shows
many peaks in the 20-30 cycles/day frequency interval which are due to
pressure (p) mode pulsations in the secondary star. Some of these
peaks are separated by 0.83 cycles/day which is the orbital
period. This implies that the amplitude of the $\delta$ Sct pulsations
are correlated with the orbital period (see Shibahashi \& Kurtz 2012
for a discussion on how power spectra can be used to measure radial
velocities in binary systems). The power spectrum of KIC 10661783
shows peaks in its power spectrum in a similar frequency range
(Southworth et al. 2011) while the p-modes seen in KIC 4544587
(Hambleton et al. 2013) are at a higher frequency range (40-50
cycles/day). We defer a full analysis of these {\kep} data for a
dedicated paper.

\subsection{KIC 7431243 (V363 Lyr): A Cataclysmic Variable}

KIC 7431243 was found to be a moderately blue source in the KIS
($g-r$=0.47, Greiss et al. 2012) and in our survey it was found to
show rapid flux changess superimposed upon an irregular variation
(Figure \ref{int-light}). KIC 7431243 matches the variable star V363
Lyr which was discovered as a Cataclysmic Variable (CV) by Hoffmeister
(1967), whilst Kato et al (2001) found that it shows outbursts of
duration 7--8 days every $\sim$21 days. There are several dozen known
CVs in the {\kep} field (see Scaringi et al. 2013 and Howell et
al. 2013).

\begin{figure*}
\begin{center}
\setlength{\unitlength}{1cm}
\begin{picture}(18,4.5)
\put(0.8,-5.5){\includegraphics{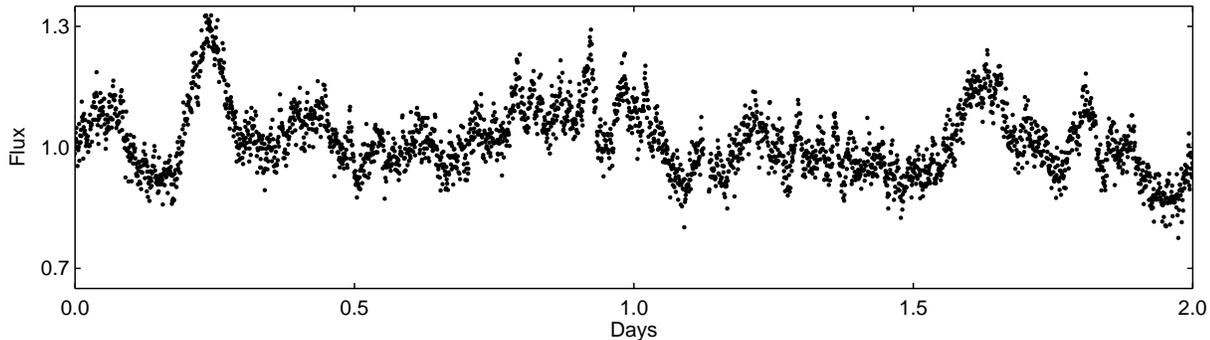}}
\end{picture}
\caption{The {\kep} short cadence light curve of KIC 7431243 (V363
  Lyr) shown over a 2 day time interval.}
\label{v363lyr-light}
\end{center}
\end{figure*}

KIC 7431243 was observed using {\kep} in Q16 for 5.2 days and (not
surprisingly) no outbursts were seen (Figure \ref{v363lyr-light}). The
power spectrum of the light curve shows peaks corresponding to 4.68
hrs and 4.47 hrs. If we attribute the longer period to the super-hump
period (super-humps are caused by the precession of the accretion
disk) and the shorter period to the orbital period, we find the
fractional excess, $\epsilon^{+}=(P_{sh}-P_{orb})/P_{orb}$=4.7
percent. Using the relationship of Patterson et al. (2005) this would
imply a mass ratio, $q=M_{2}/M_{1}\sim0.21$.

Using the secondary star mass ($M_{2}$) -- orbital period relationship
for CVs ($M_{2}=0.065 P^{5/4}_{hrs}$, Warner 1995), we find for a CV
with $P_{orb}$=4.47 hrs, $M_{2}$=0.42\Msun (0.45\Msun for
$P_{orb}$=4.68 hrs). Super-humps are thought to be restricted to
systems where the mass ratio, $q=M_{2}/M_{1}<0.33$, (see Schreiber
2007 for details). If super-humps are present in KIC 7431243 then this
may suggest that the white dwarf in this binary has a mass
$M_{1}>1.28$ \Msun assuming $M_{2}$=0.42\Msun ($M_{1}>1.36$ \Msun for
$M_{2}$=0.45\Msun).  Given the potentially high mass of the white
dwarf, we urge phase resolved optical spectroscopy this system.

\section{Conclusions}

This project set out to identify sources in the {\kep} field which
showed variability on a timescale of 1 hour or less. The most
potentially interesting of these variable sources would then have been
subject of bids to obtain {\kep} data in Short Cadence. We have
identified more than 100 strongly variable sources and we have been
succesful in obtaining {\kep} SC light curves of 18 of these sources.

Many of them are $\delta$ Scuti stars which show an astonishing range
of variability, the star with the shortest dominant period being 24
min. We also identify one $\delta$ Scuti star as being in an eclipsing
or contact binary with an orbital period of 1.21 days. As currently
only two other such systems are known in the {\kep} field, this will
provide the means to study binary evolution in more detail.  We have
also obtained {\kep} SC data of three contact binaries and one
previously known Cataclysmic Variable. The {\kep} observations of one
flare star and one pulsating DA white dwarf are reported elsewhere.

We provide a range of images and data products through the Armagh
Observatory Web Site (star.arm.ac.uk/rats-kepler). These include the
reduced images so that users can perform photometric measurements
using their favoured reduction packages. We also provide the detrended
light curves and the photometric variability parameters of each source
observed in our survey.

\section{Acknowledgements}

The Isaac Newton Telescope is operated on the island of La Palma in
the Spanish Observatorio del Roque de los Muchachos of the Instituto
de Astrofisica de Canarias (IAC) with financial support from the UK
Science and Technology Facilities Council. We would like to thank the
ING and MDM staff for their support. Observations were also made with
the Gran Telescopio Canarias (GTC) which is also sited on La Palma and
run by the IAC.  Armagh Observatory is supported by the Northern
Ireland Government through the Department of Culture, Arts and
Lesuire. We thank Wojtek Pych for the use of his difference imaging
software {\tt diapl2}. DS acknowledges support of STFC through an
Advanced Fellowship. Funding for the Stellar Astrophysics Centre
(Aarhus) is provided by The Danish National Research Foundation. The
research is supported by the ASTERISK project (ASTERoseismic
Investigations with SONG and Kepler) funded by the European Research
Council (Grant agreement no.: 267864).  We thank the referee for
helpful comments which helped to significantly improve the paper.

{}

\appendix 

\section{Tables}

\begin{table*}
\centering
\caption{List of fields observed using in the INT in 2011, field ID
  corresponds to our own internal designations, field centres
  correspond to a point in CCD 4.}
\label{field_centres}
\begin{tabular}{ c l c c c c }
\hline
Date & Field ID & RA DEC & Date(dd-mm-yy) & Field ID & RA DEC (J2000) \\
 (dd-mm-yy) & & (J2000) & (dd-mm-yy) & & (J2000) \\
\hline
 11-07-11 & 522 & 19:46:51 +49:43:35 &  02-08-11 & 382 & 19:22:42 +50:39:03 \\
 11-07-11 & 542 & 19:45:05 +47:36:21 &  02-08-11 & 408 & 19:23:33 +49:11:06 \\
 11-07-11 & 541 & 19:46:43 +47:14:21 &    03-08-11 & 46 & 19:01:23 +40:59:18 \\ 
 11-07-11 & 349 & 19:39:18 +39:16:51 &    03-08-11 & 63 & 19:05:58 +37:58:36 \\ 
 11-07-11 & 539 & 19:49:56 +46:30:21 &    03-08-11 & 260 & 19:11:38 +46:19:02 \\
 11-07-11 & 567 & 19:57:19 +43:59:07 &    03-08-11 & 73 & 19:11:46 +38:20:36 \\ 
 11-07-11 & 570 & 19:52:40 +45:05:09 &    03-08-11 & 259 & 19:13:14 +45:57:02 \\
 12-07-11 & 94 & 18:44:31 +47:36:58 &     03-08-11 & 380 & 19:14:02 +51:34:03 \\
 12-07-11 & 100 & 18:45:45 +48:09:58 &    03-08-11 & 373 & 19:14:10 +50:39:03 \\
 12-07-11 & 93 & 18:46:25 +47:14:58 &     04-08-11 & 72 & 19:02:36 +39:37:36 \\ 
 12-07-11 &  234 & 18:57:27 +49:21:54 &   04-08-11 & 236 & 19:06:53 +48:26:54 \\
 12-07-11 & 575 & 19:55:42 +45:16:09 &    04-08-11 & 264 & 19:16:22 +46:08:02 \\
 12-07-11 & 503 & 19:53:20 +42:09:20 &    04-08-11 & 296 & 19:18:56 +45:13:52 \\
 12-07-11 & 488 & 19:53:27 +40:19:20 &    04-08-11 & 277 & 19:19:29 +47:14:02 \\
 13-07-11 & 99 & 18:47:40 +47:47:58 &     04-08-11 & 269 & 19:19:33 +46:19:02 \\
 13-07-11 & 105 & 18:48:54 +48:20:58 &    04-08-11 & 406 & 19:26:55 +48:27:05 \\
 13-07-11 & 92 & 18:48:18 +46:52:58 &     05-08-11 & 245 & 19:02:50 +50:05:54 \\
 13-07-11 & 157 & 19:15:34 +42:19:19 &    05-08-11 & 76 & 19:07:31 +39:26:36 \\ 
 13-07-11 & 469 & 19:43:31 +44:27:24 &    05-08-11 & 281 & 19:20:36 +43:01:58 \\
 13-07-11 & 446 & 19:40:23 +46:44:30 &    05-08-11 & 223 & 19:28:21 +39:15:28 \\
 13-07-11 & 354 & 19:42:06 +39:27:51 &    05-08-11 & 325 & 19:36:07 +41:32:30 \\
 14-07-11 & 104 & 18:50:50 +47:58:58 &    05-08-11 & 583 & 20:06:28 +44:32:07 \\
 14-07-11 & 97 & 18:51:29 +47:03:58 &     05-08-11 & 585 & 20:03:23 +45:16:09 \\
 14-07-11 & 103 & 18:52:45 +47:36:58 &    06-08-11 & 578 & 20:01:52 +44:43:09 \\
 14-07-11 & 110 & 18:54:01 +48:09:58 &    06-08-11 & 555 & 19:59:31 +47:03:21 \\
 14-07-11 & 204 & 19:24:20 +37:36:28 &    06-08-11 & 499 & 19:59:13 +40:41:20 \\
 14-07-11 & 214 & 19:29:50 +37:58:28 &    06-08-11 & 566 & 19:58:51 +43:37:09 \\
 14-07-11 & 213 & 19:31:14 +37:36:28 &    06-08-11 & 580 & 19:58:44 +45:27:07 \\
 15-07-11 & 8 & 18:46:20 +42:51:19 &      06-08-11 & 588 & 19:58:37 +46:22:09 \\
 15-07-11 & 7 & 18:48:05 +42:29:19 &      06-08-11 & 556 & 19:57:53 +47:25:21 \\
 15-07-11 & 12 & 18:51:04 +42:40:19 &     07-08-11 & 581 & 19:57:09 +45:49:09 \\
 15-07-11 & 18 & 18:52:17 +43:13:19 &     07-08-11 & 494 & 19:54:53 +40:52:20 \\
 15-07-11 & 321 & 19:31:44 +41:43:30 &    07-08-11 & 550 & 19:54:41 +47:14:21 \\
 15-07-11 & 297 & 19:28:09 +43:56:55 &    07-08-11 & 562 & 19:54:17 +43:48:09 \\
 15-07-11 & 328 & 19:31:40 +42:38:30 &    07-08-11 & 559 & 19:52:56 +48:31:21 \\
 16-07-11 & 114 & 18:57:18 +44:04:28 &    07-08-11 & 563 & 19:52:45 +44:10:09 \\
 16-07-11 & 120 & 18:58:32 +44:37:28 &    07-08-11 & 471 & 19:52:37 +43:10:24 \\
 16-07-11 & 126 & 18:59:46 +45:10:28 &    08-08-11 & 538 & 19:51:32 +46:08:21 \\
 16-07-11 & 314 & 19:31:45 +40:48:30 &    08-08-11 & 560 & 19:51:16 +48:53:21 \\
 16-07-11 & 309 & 19:28:51 +40:37:30 &    08-08-11 & 483 & 19:50:35 +40:08:20 \\
 16-07-11 & 338 & 19:35:06 +38:32:48 &    08-08-11 & 490 & 19:50:33 +41:03:20 \\
 16-07-11 & 219 & 19:34:00 +37:47:28 &    08-08-11 & 497 & 19:50:28 +41:58:18 \\
 17-07-11 & 28 & 18:48:05 +44:52:19 &     08-08-11 & 528 & 19:50:12 +49:54:35 \\
 17-07-11 & 27 & 18:49:54 +44:30:19 &     08-08-11 & 465 & 19:49:37 +42:59:24 \\
 17-07-11 & 133 & 18:59:24 +46:05:28 &    09-08-11 & 473 & 19:49:34 +43:54:24 \\
 17-07-11 & 55 & 18:59:05 +42:16:18 &     09-08-11 & 477 & 19:49:10 +39:35:20 \\
 17-07-11 & 412 & 19:28:28 +49:00:06 &    09-08-11 & 484 & 19:49:09 +40:30:20 \\
 17-07-11 & 436 & 19:32:35 +46:44:27 &    09-08-11 & 521 & 19:48:33 +49:21:35 \\
 17-07-11 & 439 & 19:38:53 +46:11:30 &    09-08-11 & 533 & 19:48:22 +45:57:19 \\
 01-08-11 & 36 & 18:55:39 +40:37:18 &     09-08-11 & 540 & 19:48:20 +46:52:21 \\
 01-08-11 & 238 & 19:02:58 +49:10:54 &    09-08-11 & 466 & 19:48:07 +43:21:23 \\
 01-08-11 & 139 & 19:02:58 +46:16:28 &    10-08-11 & 530 & 19:46:45 +50:38:35 \\
 01-08-11 & 131 & 19:03:05 +45:21:28 &    10-08-11 & 548 & 19:46:36 +48:09:21 \\
 01-08-11 & 376 & 19:21:01 +50:06:03 &    10-08-11 & 467 & 19:46:35 +43:43:23 \\
 02-08-11 & 240 & 18:59:00 +49:54:54 &    10-08-11 & 486 & 19:46:14 +41:14:17 \\
 02-08-11 & 151 & 19:03:09 +43:25:19 &    10-08-11 & 516 & 19:45:13 +49:10:35 \\
 02-08-11 & 141 & 19:08:12 +41:24:19 &    10-08-11 & 535 & 19:45:11 +46:41:21 \\
 02-08-11 & 154 & 19:09:37 +42:52:19 &    10-08-11 & 454 & 19:45:09 +42:15:24 \\
 02-08-11 & 390 & 19:22:35 +51:34:03 & & &  \\
\hline
\end{tabular}
\end{table*}

\begin{table*}
\centering
\caption{List of fields observed using in the INT in 2012, field ID
  corresponds to our own internal designations, field centres
  correspond to a point in CCD 4.}
\label{field_centres}
\begin{tabular}{ c l c c l c }
\hline
Date & Field ID & RA DEC & Date & Field ID & RA DEC (J2000) \\
 (dd-mm-yy) & & (J2000) & (dd-mm-yy) & & (J2000) \\
\hline
03-08-12 & 5 & 18:39:52 +43:24:19     & 08-08-12 & 1010 & 19:20:19 +43:38:30   \\
03-08-12 & 67 & 18:59:22 +39:26:36      & 08-08-12 & 203 & 19:25:43 +37:14:28         \\  
03-08-12 & 144 & 19:03:15 +42:30:19    & 08-08-12 & 251 &  19:06:41 +50:16:54          \\
03-08-12 & 300 & 19:23:31 +45:02:56    & 08-08-12 & 289 &  19:19:02 +44:18:58          \\
05-08-12 & 463 & 19:42:05 +43:54:24    & 08-08-12 & 383 &   19:20:57 +51:01:03         \\
05-08-12 & 1001 & 18:45:00 +47:21:36   & 09-08-12 & 1005 &  19:12:34 +43:30:14        \\
05-08-12 & 1007 & 18:59:02 +48:42:37   & 09-08-12 & 1006 &  19:17:19 +39:27:18        \\
05-08-12 & 155 & 19:08:06 +43:14:17    & 10-08-12 & 1006 &  19:17:19 +39:27:18        \\
05-08-12 & 568 & 19:55:46 +44:21:07    & 10-08-12 & 1008 &  19:11:33 +45:43:44        \\
05-08-12 & 463 & 19:42:05 +43:54:24   & 10-08-12 & 210 & 19:25:41 +38:09:28          \\
05-08-12 & 546 & 19:49:53 +47:25:22   & 10-08-12 & 222 & 19:29:47 +38:53:28           \\
05-08-12 & 523 & 19:45:08 +50:05:35   & 10-08-12 & 365 & 19:15:57 +49:22:03           \\
06-08-12 & 1 & 18:46:52 +41:56:18     & 10-08-12 & 368 &  19:10:48 +50:28:03          \\
06-08-12 & 1002 & 19:04.62 +42:45:48  & 10-08-12 & 374 &   19:12:25 +51:01:03         \\
06-08-12 & 168 & 19:10:56 +44:20:19   & 11-08-12 & 1009 & 19:09:59 +47:17:07         \\
06-08-12 & 175 & 19:17:49 +39:37:14   & 11-08-12 & 1011 &  19:18:30 +45:33:11        \\
06-08-12 & 181 & 19:19:14 +40:10:15   & 11-08-12 & 1012 &  19:19:12 +49:57:51        \\
06-08-12 & 305 & 19:28:07 +44:52:02   & 11-08-12 & 292 & 19:25:07 +43:45:58           \\
06-08-12 & 342 & 19:39:19 +38:21:51   & 11-08-12 & 359 & 19:46:21 +39:16:51           \\
07-08-12 & 1013 & 19:29:12 +50:19:02  & 11-08-12 & 392 &  19:19:00 +52:18:03          \\
07-08-12 & 226 & 19:00:03 +48:04:54    & 11-08-12 & 401 & 19:23:41 +48:16:06           \\
07-08-12 & 239 & 19:00:59 +49:32:54    & 12-08-12 & 10 &   18:42:48 +43:35:20          \\
07-08-12 & 311 & 19:25:55 +41:21:30    & 12-08-12 & 2  &  18:45:08 +42:18:20          \\
07-08-12 & 327 & 19:33:09 +42:16:30    & 12-08-12 & 3 &  18:43:23 +42:40:20           \\
07-08-12 & 332 & 19:37:34 +42:05:30    & 12-08-12 & 4 &  18:41:38 +43:02:20            \\
07-08-12 & 351 & 19:36:26 +40:00:51    & 12-08-12 & 496 &  19:51:57 +41:36:20          \\
08-08-12 & 1003 &  19:06:31 +43:54:48 & 12-08-12 & 6 &  18:49:50 +42:07:20            \\
08-08-12 & 1004 &  19:19:55 +42:47:29 & 12-08-12 & 9 &  18:44:35 +43:13:20            \\
\hline
\end{tabular}
\end{table*}

\begin{table*}
\centering
\caption{List of fields observed using the MDM 1.3m telescope in 2012, field ID
  corresponds to our own internal designations, field centres
  correspond to a point approximately in the centre of the chip.}
\label{field_centres}
\begin{tabular}{ c l c c l c }
\hline
Date & Field ID & RA DEC & Date  & Field ID & RA DEC\\
 (dd-mm-yy) & & (J2000) & (dd-mm-yy) & & (J2000) \\
\hline
16-05-2012 & 81 & 19:12:07   +39:16:50 &       21-05-2012 & 174 & 19:19:15 +39:15:15 \\
16-05-2012 & 82 & 19:10:42   +39:38:50  &      21-05-2012 & 175 & 19:17:49 +39:37:15  \\ 
16-05-2012 & 83 & 19:09:16   +40:00:51 &       21-05-2012 & 226 &  19:00:03 +48:04:55 \\
16-05-2012 & 84 & 19:07:49   +40:22:55 &       21-05-2012 & 351 & 19:36:26 +40:00:51   \\
17-05-2012 & 85 & 18:49:13   +45:58:48 &       22-05-2012 & 212 & 19:22:51 +38:53:28   \\
19-05-2012 & 135 & 19:10:10   +44:50:04  &     22-05-2012 & 218 & 19:24:11 +39:26:28   \\
19-05-2012 & 142 & 19:07:08   +41:47:26  &     22-05-2012 & 273 & 19:13:03 +47:47:02  \\
19-05-2012 & 149 & 19:07:04   +42:42:34 &      22-05-2012 & 95 & 18:42:35 +47:58:58   \\
19-05-2012 & 89 & 18:41:41   +47:26:48 &       23-05-2012 & 106 & 18:46:58 +48:42:58   \\
20-05-2012 & 122 & 18:54:54 +45:21:28 &   23-05-2012 & 286 & 19:23:37 +43:12:58  \\
20-05-2012 & 160 & 19:11:03 +43:25:19 &   23-05-2012 & 288 & 19:20:34 +43:56:58 \\
20-05-2012 & 161 & 19:09:32 +43:47:19 &   23-05-2012 & 329 & 19:30:09 +43:00:30 \\
20-05-2012 & 162 & 19:08:00 +44:09:19 & & & \\
\hline
\end{tabular}
\end{table*}

\begin{table*}
\caption{The full set of parameters which are included in our data products.}
\label{fits-files-parameters}
\begin{center}
\begin{tabular}{ll}
\hline
Parameter & Notes \\
KIC\_ID & The {\kep} Input Catalog (Brown et al. 2011) star number;\\
RA, DEC & Right Ascension and Declination (J2000);\\
$g_{mag}$ & Taken from the KIS, Greiss et al. (2012a,b);\\
$g_{err}$ &  Taken from the KIS Greiss et al. (2012a,b);.\\
U\_g, g\_r & $(U-g)$, $(g-r)$ taken from Greiss et al. (2012a,b);\\
LS\_Period & The period of the most prominent peak in the Lomb Scargle periodogram in days and LS\_Period\_Mins (mins);\\
Log10\_LS\_Prob & The False Alarm Probability of the most prominent period in the Lomb Scargle periodogram 
  (in units of log 10);\\
Alarm & The alarm variability statistic (Tamuz, Mazeh, and North 2006);\\
AoV\_Period & The Period determined from the AoV test in days and AOV\_Period\_Mins (mins);\\
AoV & The AoV variability statistic Schwarzenberg-Czerny (1989, 1996);\\
AOV\_SNR & the S/N ratio of the peak measured over the full periodogram;\\
AOV\_NEG\_LN\_FAP & The negative of the natural logarithm of the formal false alarm probability;\\
Chi2 & The reduced $\chi^{2}$ value of the light curve tested against the constant mean value (with 5$\sigma$ clipping);\\
StdDev & the standard deviation (root mean squared) of the light curve;\\
Field & Our internal naming convention for the field pointing;\\
Chip & For the INT/WFC the chip id. There was only one chip for the MDM observations;\\
FieldChip & The Field-Chip Combination\\
ID & Our internal naming convention for the source. A 6 digit number implies the light curve was derived; \\ 
   & using {\tt diapl}, while for those derived using {\tt sextractor}, the numbering system starts from 1;\\
X,Y & The X,Y coordinates of the source on the chip;\\
grats, g\_{r}\_{rats} &The $g$ mag and $(g-r)$ colour at the time of our observations;\\
Flag & `0'  Variability on ls\_{per}\_{min}; `1' Probable variability on ls\_{per}\_{min}; `2' Clear long timescale high amplitude variable;\\
     & `3'  Not variable on ls\_{per}\_{min}; `4'  Variable on period other than  ls\_{per}\_{min}; `5' Possibile variability in genera;l\\
     & `6'  not likely to be variable; `7' bad light curve; `8' Eclipse; `9' Possible eclipse;\\
     & `10' Variability likely due to systematic trend; `11' Known bad columns on chip;\\
     & `12' Apparent long period could be due to residual systematic trends; `13' Image shows close stellar companion;\\
Tstart, Tstop & The start and end date in MJD of the sequence of $g$ band observations.\\
medFAP & The median log FAP for the chip which the source is located\\ 
\hline
\end{tabular}
\end{center}
\end{table*}

\setcounter{table}{4}
\landscape
\begin{table}
\caption{Table showing all of our variable stars selected using
  '$n$=18' and which passed our manual verification phase (see \S 3.2)
  along with a selection of parameters. }
\label{variables-list}
\begin{center}
\begin{tabular}{rccrrrrrrrrrr}
\hline
 KIC & RA & DEC & $g_{KIS}$ & $(U-g)_{KIS}$ & $(g-r)_{KIS}$ & $g_{rats}$ & $(g-r)_{rats}$ &
Alarm & AOV Period& AOV & LS\_Per & LS\_Log\_FAP\\
ID & J2000 & J2000 & mag & mag & mag & mag & mag & & mins & & mins & \\
\hline
11911480&19:20:24.91&+50:17:22.4&18.09   & -0.39 & 0.06 & 18.13 &-0.05 & 0.02 & 4.89  &14.75  & 4.9  &-6.93\\    
8293193&19:17:55.25&+44:13:26.1&18.42    & -0.31 & 0.08 & 18.41 & 0.00 & 1.88 & 5.17  &5.81   & 5.2  & -4.46\\   
       &19:29:00.80&+44:56:59.2&         &       &      & 13.95 &      &  1.43& 9.93  & 9.29  & 10.1 & -5.51 \\  
12647528&19:22:50.90&+51:45:31.5&14.31   &  1.73 & 1.05 &14.35 & 0.99 &2.75  &  43.4  & 10.36 & 10.8 & -4.83\\   
10728590&19:23:18.49&+48:02:09.0& 19.14  &       & 0.97 & 19.37 & 1.01 & 4.25 & 13.9  & 8.73  & 13.8 &-4.81\\    
10936077&19:52:53.44&+48:19:35.6& 15.85  & 0.70  & 0.70 & 15.90 & 0.56 & 2.62 & 31.8  & 12.19 & 16.2 & -4.80\\   
7356523&19:19:30.70&+42:58:08.5&19.20    & 0.94  & 0.65 & 19.20 & 0.65 & 2.00 & 18.8  & 6.91  & 18.5 & -4.44\\   
9899481&19:41:00.84&+46:44:58.3&19.66    & 0.49  & 0.64 & 19.61 & 0.53 & 2.99 & 36.4  & 10.77 & 19.3 & -5.11\\   
6665002&18:46:27.78&+42:10:34.9&19.50    & 0.13  & 0.68 & 19.45 & 0.64 & 3.34 & 21.0  & 9.38  & 21.2 & -4.66\\   
8123702&19:57:39.81&+43:55:07.7&13.46    & 2.04  & 1.17 & 13.50 & 1.11 & 4.54 & 64.9  & 8.78 & 21.8 & -3.98\\   
       &18:55:13.43&+43:57:31.6&19.75    &       &      & 19.61  & 1.08 & 2.21& 22.9  & 8.10  & 23.2 & -3.83 \\  
6109859&19:07:56.19&+41:26:33.1&16.19    & 0.35  & 0.17 & 16.19 & 0.17 & 8.22 & 23.1  & 39.38 & 23.4 & -11.64\\  
3223460&19:12:32.15&+38:23:00.1&13.71    & 0.27  & 0.25 & 13.74 & 0.16 & 7.58 & 24.4  & 29.58 & 24.0 & -10.71\\  
       &19:53:11.03&+48:39:39.1&         &       &       &14.79 & 0.63 & 4.06 & 44.6  & 14.76 & 25.9 & -5.42\\   
       &19:18:03.08&+39:26:21.8&         &       &       &       &       &5.28& 26.4  & 16.88 & 26.2 & -8.03\\   
11360026&19:45:03.62&+49:06:01.0&15.01  & 0.24 & 0.29 & 14.99 & 0.16 & 9.25   & 32.9  & 18.47 & 32.5 &-9.59\\    
9813390&18:49:08.86&+46:40:04.2&16.86   & 1.30&1.14&16.98&1.04&10.29          & 66.2  & 24.35 &34.7  &-5.17\\    
10031075&19:54:21.16&+46:57:39.9&15.17  &0.63&0.76&15.18&0.64&2.71            & 38.2  & 9.62  &35.1  &-5.57\\    
8118210&19:52:18.66&+43:58:11.8&17.02   &0.46&0.47&17.09&0.33&3.98            & 65.4  & 12.07 &35.4  &-6.31\\    
7960631&19:28:05.53&+43:45:45.5&15.31   &0.30&0.31&15.32&0.18&7.28            & 63.5  & 23.15 &35.6  &-7.96\\    
9479634&19:48:39.16&+46:03:47.1&14.65   &0.39&0.32&14.63&0.20&11.23           & 36.0  & 51.00 &36.4  &-12.4\\    
5772488&19:00:07.98&+41:02:51.8&17.67   &0.21&0.33&17.70&0.16&9.54            & 65.5  & 22.88 &37.3  &-8.15\\    
11723564&19:47:11.37&+49:53:13.7&12.81  &0.38&0.40&12.74&0.17&7.47            & 66.4  & 18.83 &38.0  &-7.91\\    
10253681&18:46:40.86&+47:18:28.1&17.22   &0.60&0.69&17.04&0.63&5.61           & 37.6  & 25.38 &39.6  &-8.51\\    
9786930&19:51:01.55&+46:34:24.6&19.00   &1.29&1.00&19.03&0.77&8.33            & 47.3  & 18.77 &41.1  &-6.77\\    
10975348&19:26:46.09&+48:25:30.9&18.88  &0.19&0.34&18.72&0.07&13.32           & 54.0  & 32.84 &41.2  &-9.59\\    
9364179&19:56:24.52&+45:48:24.1&14.38   &0.44&0.43&14.42&0.36&6.93            & 43.4  & 17.11 &41.2  &-7.51\\    
9294308&19:47:09.39&+45:44:32.1&13.66   &0.25&0.34&13.64&0.19&6.09            & 44.7  & 17.89 &42.9  &-8.05\\    
7698266&19:46:50.79&+43:19:02.3&13.22   &0.21&0.41&13.26&0.32&6.54            & 43.7  & 18.00 &43.6  &-7.86\\    
10353926&19:47:53.67&+47:26:56.3&14.89  &0.22&0.28&14.90&0.31&9.32            & 45.0  & 26.15 &44.1  &-9.90\\    
7625723&19:47:11.80&+43:16:37.4&14.18   &1.42&1.02&14.21&0.95&8.22            & 48.7  & 19.43 &44.2  &-7.70\\    
9417741&19:47:39.58&+45:55:07.6&14.72   &0.29&0.28&14.78&0.20&15.78           & 45.3  & 75.25 &44.4  &-13.3\\    
\hline
\end{tabular}
\end{center}
\end{table}
\endlandscape

\setcounter{table}{4}
\landscape
\begin{table}
\caption{Table showing all of our variable stars selected using
  '$n$=18' and which passed our manual verification phase (see \S 3.2)
  along with a selection of parameters.}
\label{variables-list}
\begin{center}
\begin{tabular}{rccrrrrrrrrrr}
\hline
 KIC & RA & DEC & $g_{KIS}$ & $(U-g)_{KIS}$ & $(g-r)_{KIS}$ & $g_{rats}$ & $(g-r)_{rats}$ &
Alarm & AOV Period& AOV & LS\_Per & LS\_Log\_FAP\\
ID & J2000 & J2000 & mag & mag & mag & mag & mag & & mins & & mins & \\
\hline
4262791&19:26:48.06&+39:20:35.7&15.74&4.82&0.26&15.71 & 0.21 &13.49&62.9 & 23.15 & 44.5 & -7.60\\ 
12406812&19:23:33.80&+51:17:58.9&17.24&0.17&0.36&17.24& 0.21 &14.47&46.8 & 47.66 & 46.5 & -12.06\\
7548311&19:48:28.48&+43:06:12.3&14.55&0.47&0.48 &14.52& 0.34 &10.56&66.5 & 32.02 & 46.7 & -9.56\\ 
5623923&19:32:01.53&+40:51:16.8&16.61&0.23&0.27 &16.57&0.15  &10.30&45.3 &23.26&47.4&-10.24\\     
6418095&18:44:56.53&+41:50:28.1&13.20&0.12&0.23 &13.14&0.23  &10.41&47.4 &35.16&48.3&-11.61\\     
7770746&19:49:00.55&+43:26:06.4&18.84&1.41&1.16 &19.20&1.35  &17.02&47.6 &72.20&48.5&-13.46\\     
9640005&19:09:46.28&+46:20:04.1&18.39&0.15&0.21 &18.39&0.15  &18.18&63.9 &50.71&48.5&-11.27\\     
9786165&19:50:10.98&+46:34:40.8&17.66&0.81&0.76 &18.39&1.12  &16.50&47.3 &47.53&49.7&-10.19\\     
5474065&19:53:02.53&+40:40:34.6&18.77&1.93&1.43 &18.76&1.39  &12.75&51.3 &105.44&49.9&-6.32\\     
9672731&19:59:30.69&+46:23:06.7&17.47&0.97&0.98 &17.95&1.16  &8.185&66.2 &18.77&51.1&-8.65\\      
8840638&19:55:35.07&+45:04:46.0&14.63&0.52&0.54 &14.71&0.42  &22.33&49.8 &98.99&51.6&-14.54\\     
9720306&19:43:03.19&+46:25:56.0&15.45&0.28&0.33 &15.50&0.23  &15.80&49.7 &30.96&51.9&-10.27\\     
6029053&19:08:02.00&+41:22:12.6&17.51&3.21&1.43 &17.51&1.43  &13.82&55.1 &76.53&52.9&-7.61\\      
8117771&19:51:52.82&+43:55:00.1&13.24&0.26&0.36 &13.22&0.25  &17.99&54.2 &59.45&53.4&-12.18\\     
8387281&19:54:08.13&+44:22:16.5&15.65&1.65&1.47 &15.64&1.46  & 7.73&45.5 & 49.45&55.6&-6.74\\     
4389023&19:47:47.78&+39:29:41.5&13.63&0.21&0.36 &13.64&0.25  &18.31&52.3 &59.75&56.8&-12.35\\     
5818101&19:54:18.14&+41:04:07.1&17.52&0.25&0.39 &17.48&0.28  &17.61&65.1 &32.50&57.4&-11.32\\     
7839261&19:48:18.90&+43:34:22.1&14.77&0.39&0.40 &14.77&0.31  &22.00&57.2 &92.79&58.0&-13.31\\     
8118471&19:52:35.03&+43:56:42.3&15.72&0.45&0.56 &15.75&0.46  &12.94&60.8 &28.50&59.2&-10.42\\     
5179693&19:18:32.80&+40:22:51.3&17.60&0.17&0.32 &17.74&0.29  &19.35&65.8 &64.82&61.3&-12.62\\     
7768746&19:47:00.08&+43:28:22.2&13.23&0.38&0.58 &13.18&      &18.11&54.1 &78.36  &61.5&-12.69\\   
3350736&19:34:23.34&+38:24:30.5&15.47&0.30&0.45 &15.53&0.33  &23.04&64.3 &41.69&62.1&-11.50\\     
10284901&19:43:46.41&+47:20:32.8&15.72&0.06&0.31 &15.59&0.11 &19.40&64.4 &161.97&62.1&-13.80\\    
8255272&19:53:55.44&+44:11:49.4&13.24&0.28&0.41 &13.18&0.44  &17.86&62.1 &49.52&63.0&-11.85\\     
9364721&19:57:04.39&+45:48:41.1&19.06&0.60&0.85 &19.06&0.71  &10.03&64.2 &19.32&63.7&-8.99\\      
8908767&19:56:07.19&+45:08:50.8&14.30&0.56&0.55 &14.35&0.46  &21.59&66.4 &118.8&64.0&-14.48\\     
9905251&19:48:41.25&+46:43:33.9&16.11&0.30&0.35 &16.13&0.18  &17.04&55.9 &40.88&64.4&-12.30\\     
9812716&18:47:14.89&+46:36:38.4&13.95&0.28&0.29 &13.91&0.14  &13.67&52.7 &38.82&64.9&-11.91\\     
&19:16:43.60&+39:31:27.8&&&&&                                &17.24&62.2 &61.85&65.2&-12.21\\     
8254486&19:53:11.74&+44:07:43.2&14.76&0.42&0.58&14.59&0.48   &17.17&56.6 &47.02&65.2&-12.13\\    
4916020&19:17:58.54&+40:04:54.6&14.57&0.33&0.31&14.53&0.28   &13.69&61.9 &47.96&65.8&-11.79\\    
4149801&19:18:07.97&+39:15:42.0&19.51&1.56&1.21&19.74&2.04   &18.62&54.2 &58.62&66.1&-11.72\\     
8561192&19:29:05.17&+44:41:50.8&16.43&0.34&0.49&16.40&0.48   &17.05&66.5 &62.97&66.5&-10.86\\     
\hline
\end{tabular}
\end{center}
\end{table}
\endlandscape

\end{document}